\documentclass[12pt]{iopart}
\pdfoutput=1
\usepackage{iopams} 
\usepackage{xcolor}
\usepackage{graphicx}

\usepackage[numbers]{natbib}
\usepackage{pifont}
\usepackage[style=base]{caption}
\usepackage{subcaption}
\usepackage{longtable}
\usepackage[normalem]{ulem}

\newcommand{\cmark}{\ding{51}}%
\newcommand{\xmark}{\ding{55}}%
\newcommand{\bx}{\boldsymbol{x}}

\begin{document}

\title[Linearity Characterization and Uncertainty Quantification for Spectroradiometers]{Linearity Characterization and Uncertainty Quantification of Spectroradiometers via Maximum Likelihood and the Non-parametric Bootstrap}

\author{Adam L. Pintar$^1$, Zachary H. Levine$^2$, Howard W. Yoon$^3$, Stephen E. Maxwell$^3$}

\address{$^1$ Statistical Engineering Division, National Institute of Standards and Technology, Gaithersburg, Maryland 20899-8980 USA; apintar@nist.gov}
\address{$^2$ Quantum Measurement Division, National Institute of Standards and Technology, Gaithersburg, Maryland 20899-8441 USA; zlevine@nist.gov}
\address{$^3$ Sensor Science Division, National Institute of Standards and Technology, Gaithersburg, Maryland 20899-8441 USA; smaxwell@nist.gov}

\begin{abstract}
A technique for characterizing and correcting the linearity of radiometric instruments is known by the names the ``flux-addition method'' and the ``combinatorial technique''. In this paper, we develop a rigorous uncertainty quantification method for use with this technique and illustrate its use with both synthetic data and experimental data from a ``beam conjoiner'' instrument. We present a probabilistic model that relates the instrument readout to a set of unknown fluxes via a set of polynomial coefficients. Maximum likelihood estimates (MLEs) of the unknown fluxes and polynomial coefficients are recommended, while a non-parametric bootstrap algorithm enables uncertainty quantification (e.g., it can return standard errors). 

The synthetic data represent plausible outputs of a radiometric instrument and enable testing and validation of the method. The MLEs for these data are found to be approximately unbiased, and confidence intervals derived from the bootstrap replicates are found to be consistent with their target coverage of 95~\%.  For the polynomial coefficients, the relative bias was less than 1~\% and the observed coverages range from 90~\% to 97~\%.  The experimental data set is used to illustrate how a complete calibration with uncertainties can be achieved using the method plus one well-known flux level. The uncertainty contribution attributable to estimation of the instrument's nonlinear response is less than 0.02~\% over most of its range.
\end{abstract}
 
\noindent{\it Keywords}: Satellite Based Measurements, Spectroradiometer, Calibration, Maximum Likelihood, Bootstrap

\section{Introduction}
\label{sec:intro}

Indicating instruments, such as spectroradiometers, pressure gauges, or scales, require calibrations so that their output may be considered a measurement result \cite{vim2012}. However, such instruments are often calibrated using only a small number of known stimuli, while the subsequent calibrations are expected to be valid over a dynamic range that can span several orders of magnitude. If an instrument's response is known to be perfectly linear in the stimulus, only a single known non-zero stimulus, combined with estimates of uncertainty, is required for calibration. Thus, assessing and correcting the linearity of an instrument's response is a critical component of instrument calibration.

One method for assessing linearity is the ``flux-addition'' method or the ``combinatorial technique.'' The method is used in the calibration of many types of instruments~\cite{white_general_2008}.  The technique exploits the property of a linear function $f(a+b) = f(a) + f(b)$ to construct a system of equations that enables the recovery of the response function of an indicating instrument, up to an overall scaling factor, using a set of unknown stimuli.

A technique for recovering the response function involves assuming a polynomial form and setting up a system of equations that includes both the coefficients of the polynomial and the unknown stimuli. This system can be solved using either pseudoinverse methods or numerical optimization routines. However, because both the stimuli and the response function are unknown, additional constraints are needed to obtain a unique solution. 
We know of no published standard method to recover rigorous uncertainty statements, such as confidence intervals, about the estimated parameters. While repeating the experiment until a sufficient number of estimates of the parameters are generated to characterize the uncertainty to the desired level would work, it is not an efficient use of resources.  Instead, here we develop a maximum likelihood inversion technique that, in concert with a bootstrap algorithm~\cite{efron1993introduction}, may be used as part of a full uncertainty analysis when calibrating an indicating instrument. This algorithm is a novel contribution of this paper. We then apply this technique to two systems. The first system is synthetic data based on simulations of a light source and radiometer similar to that used in calibrations of Earth observing satellites, and the second system is NIST's optical Beam Conjoiner~\cite{thompson_beamcon_1994}.

In the simulation, we make a direct comparison of the recovered parameters to the known inputs, enabling an evaluation of the algorithm's performance. We also explain how in the case of calibrations of imagers, a linear calibration may be achieved even in the presence of non-uniform illumination, a critical problem in the calibration of remote sensing instruments, which may have many pixels that observe different parts of a calibration standard.  In the application to the Beam Conjoiner, we compare the results of the new method to the results obtained using the traditional inversion technique.  We also show how uncertainty in the estimated nonlinear detector response can be incorporated into a calibration using the output from the bootstrap algorithm. 

The remainder of the article is organized as follows: Section~\ref{sec:methodology} describes the probabilistic model and techniques used to estimate the fluxes and polynomial coefficients as well as quantify uncertainty in those estimates.
Section~\ref{sec:sim_study} presents the results of a simulation study using realistic assumptions based on NIST's experience calibrating earth observing satellites.
Section~\ref{sec:beamcon_results} presents results for an experimental data set collected at NIST using the Beam Conjoiner and a commercial spectroradiometer.
Finally, Section~\ref{sec:summary} summarizes the main contributions of the article.

\section{Methods}\label{sec:methodology}

Consider an indicating instrument output $n$ in response to a stimulus $\Phi$.
In the absence of noise, an instrument has an ideal linear response when $n\propto\Phi$.
We are concerned in this paper with instruments that can be described by a relation $n=h(\Phi)$, with $h$ some well-behaved, monotonic, and continuous function.  In this case, it is of interest to find a ``linearizing function'' $h^{-1}$ so that $h^{-1}(n) = \Phi$. This is a ``linearizing function'' in the sense that its output is proportional to the stimulus.  For instruments with responses of the type considered here, the linearizing function $h^{-1}$ can be approximated by a polynomial
\begin{equation*}
h^{-1}(n) = \beta_0 + \sum_{m = 1}^{p} \beta_m n^m,
\end{equation*}
with $p$ chosen to achieve the desired accuracy. 
Other functional forms, such as splines, may be substituted into the method presented here, but for concreteness we restrict the development to polynomial functions.

To use a single known non-zero stimulus to calibrate an indicating instrument, a first step is to estimate the shape of the linearizing function $h^{-1}$. We reduce this problem to estimating the above coefficients, up to an unknown scale factor, following the lead of Ref.~\cite{white_general_2008}.  The method is variously called the combinatorial technique, or in the case of radiometric calibrations, the flux-addition or Beam Conjoiner method. No prior knowledge of the magnitude of the stimuli is required, although they must be stable, repeatable, and chosen to sample the dynamic range of the instrument. The technique enables estimation of the magnitude of the observed stimuli and the coefficients of the the linearizing function, up to the aforementioned overall scale factor.  Next, a bootstrap algorithm  enables quantification of uncertainty in the parameters. Finally, measurement of a known stimulus then constrains the overall scaling of all measured parameters. 

In the method presented here, all three steps originate from a generative probabilistic model.  The model is generative in the sense that it could be used to generate data that mimic those from a real-world experiment.  It is probabilistic in the sense that such data would be generated from a joint probability distribution, not deterministically.  The model is described next.

\subsection{Statistical Model}
\label{sec:stat_model}

For concreteness, we develop our model in the context of calibration of a radiometric measurement device that is viewing a lamp illuminated integrating sphere with a set of $J>2$ lamps that can be switched on and off.  This choice was made because of an interest in optical satellite calibrations, but it does not restrict generalization of the model to other systems.  Integrating spheres with two variable light sources have been considered for linearity characterization before \cite{shin2013high}. Variable light sources can easily be accommodated in our model by considering each specified level to be a separate lamp and restricting the allowed configurations such that one light source cannot have more than one level on at once. An additional method, which we describe after this simpler version of the model, allows the introduction of correlations between these levels, as might be the case when using a variable aperture.
The first model is described by the relations
\begin{eqnarray}
  \label{eq:stat_model_prob}
  n_i ~ & \sim ~\mbox{Normal}\left(\alpha_0 + \sum_{m=1}^p \alpha_m P_m\left(s(\Phi_i)\right),~\sigma^2\right) \nonumber \\
  \Phi_i &= \sum_j x_{ij}\phi_j \nonumber \\
  \sum_j \phi_j ~ & \sim ~ \mbox{Normal}\left(\Phi_{\rm max}, \tau^2\right) \nonumber \\
  \alpha_1 ~  & \sim ~ \mbox{Normal}\left(\frac{\Phi_{\rm max}}{2}, \gamma^2\right) \\
  \alpha_m ~  & \sim ~ \mbox{Normal}\left(0, \gamma^2\right) \mbox{ for } m > 1 \nonumber \\
  \gamma~ & \sim ~ \mbox{Exponential}\left(\lambda\right),\nonumber
\end{eqnarray}
where $\sim$ indicates that the quantity on the left follows the distribution on the right, $\mbox{Normal}\left(\mu,\xi^2\right)$ is the normal distribution with mean $\mu$ and standard deviation $\xi$, $\mbox{Exponential}\left(\lambda\right)$ is the exponential distribution with mean $\lambda$, $n_i$ is the instrument reading from data point $i$, $\Phi_i$ is the stimulus (here a flux) for data point $i$, $s$ is a linear transformation of $\Phi_i$ to the interval $[-1, 1]$, $P_m$ is the Legendre polynomial of order $m$~\cite{LegendreFunction}, $\balpha = (\alpha_0, \dots, \alpha_p)$ is a vector of polynomial coefficients, $\phi_j$ is the flux of lamp $j$, $x_{ij}$ takes the values 0 or~1 for lamp $j$ being off or on in data point $i$ (and the values are restricted to only physically allowed configurations in the case of variable light sources), $\sigma$ represents noise, $\Phi_{\rm max}$ estimates the maximum flux, $\tau$ is the uncertainty in the estimate of the maximum flux, $\gamma$ is a penalty parameter that pulls the regression coefficients towards expected values \textit{a priori}, and $\lambda$ is another penalty parameter that disallows very large values of $\gamma$. 
The noise represented by $\sigma$ may be due both to instrumental sources and noise in the stimulus. The $n_i$, the $x_{ij}$, $\Phi_{\rm max}$, $\tau$, and $\lambda$ are taken to be fixed quantities and are inputs to the model.  All other parameters, i.e., $\bphi = (\phi_1, \dots, \phi_J)$, $\balpha$, $\gamma$, and $\sigma$ are estimated by maximum likelihood and those estimates are called maximum likelihood estimates (MLEs)\footnote{See for example Chapters 4 and 5 of \cite{wood2015core}}.  If the parameter $\sigma$ were known or estimated from a separate experiment, it could be treated as a known fixed quantity. The assumption of normally distributed homogeneous noise is not a unique possible choice, and the exact noise behavior of the model must be tuned to match the system under study (see Section~\ref{sec:beam_conjoiner} for an example).

The constraint $\sum_j \phi_j\sim\mbox{Normal}\left(\Phi_{\rm max},\tau^2\right)$, 
represents a measurement with all lamps on, and sets a scale for the solution.  The constraints $\alpha_1\sim\mbox{Normal}\left(\frac{\Phi_{\rm max}}{2}, \gamma^2\right)$, and $\alpha_m\sim\mbox{Normal}\left(0, \gamma^2\right) \mbox{ for } m > 1$ are required to decouple the lamp fluxes from the nonlinear response of the detector.  Without these constraints, the estimation problem is ill-posed. The constraints force the linear coefficient $\alpha_1$ toward $\frac{\Phi_{\rm max}}{2}$ because the light levels $\Phi_i$ then lie in the interval $[0, \Phi_{\rm max}]$, and over that interval, the linear coefficient is expected to be near one for the well-behaved instruments considered here. Since Legendre polynomials are orthogonal in the interval $[-1, 1]$, the interval $[0, \Phi_{\rm max}]$ is shifted and scaled to the interval $[-1, 1]$, and a slope of one on the original interval is a slope of $\frac{\Phi_{\rm max}}{2}$ on the new interval.  The higher order coefficients are forced toward zero by our imposed constraints, due to our restriction to well-behaved instruments. Note that the $n_i$ are described in terms of a forward model that is polynomial in the fluxes; a calibration is expected to result in coefficients that give the fluxes in terms of a polynomial in the $n_i$. This is discussed further in Section \ref{sec:linearization}

The strength of the forcing of the polynomial coefficients, to balance prediction variance and bias, is a parameter that is optimized within the model and thus not arbitrary.  
This forcing is related to the technique known as ridge regression, introduced originally in \cite{hoerl1970ridge}, and the technique known as LASSO (for ``Least Absolute Shrinkage and Selection Operator''), originally introduced in \cite{tibshirani1996regression}. 
Since the Gaussian distribution is leveraged to achieve the shrinkage, the similarity to ridge regression is more striking.  It also resembles standard Bayesian treatments of regression problems, see for example Section~14.8 of reference \cite{gelman2013bayesian}, which uses Gaussian prior distributions for regression coefficients, and \cite{park2008bayesian} which uses Laplace prior distributions.

The statistical model assumes stability and repeatability of lamp flux levels over the duration of the acquisition of the $n_i$.  That assumption is reflected in Equation~(\ref{eq:stat_model_prob}) by the fact that the lamp fluxes, the $\phi_j$, do not change across a data set (e.g., they don't depend on $i$). However, integrating spheres have variable throughput that depends on the average reflectance of the interior surface, and light sources change over time. In Section \ref{sec:sim_study}, the effects of variable throughput on the MLEs  are examined, and we incorporate changing light sources into our synthetic data to investigate how this affects the utility of the method.

Up to constant terms, the log-likelihood function corresponding to
Equation~(\ref{eq:stat_model_prob}) is
\begin{eqnarray}
    \label{eq:stat_model_like}
    \ell(\bPhi, \balpha, \gamma, \sigma) = 
    -\frac{1}{2\sigma^2}&\left[\sum_{i=1}^N\left(n_i - \alpha_0 - \sum_{m=1}^p\alpha_mP_m(s(\Phi_i))\right)^2\right] - \nonumber \\ 
    &N\log\sigma -\frac{1}{2\tau^2}\left(\sum_{j=1}^J\phi_j - \Phi_{\rm max}\right)^2 - \nonumber \\
    &\frac{1}{2\gamma^2}\left(\alpha_1 - \frac{\Phi_{\rm max}}{2}\right)^2 - \log\gamma - \nonumber \\
    &\frac{1}{2\gamma^2}\left[\sum_{m=2}^p\left(\alpha_m\right)^2\right] - (p-1)\log\gamma -
    \lambda\gamma .
\end{eqnarray}
Equation~(\ref{eq:stat_model_like}) is based on a joint probability distribution over all of the variables.  Hence, by definition, it is a generative model.

The integrating sphere used calibrate the Orbiting Carbon Observatory-2 \cite{rosenberg_preflight_2017} has 10 lamps, and a single lamp with a variable aperture. To make our model more like this, we make one of the lamps a variable aperture. Under the assumption that there are $J$ lamps with lamp $J$ having a variable aperture, we modify $\Phi_i$ from Equation~(\ref{eq:stat_model_prob}) to
\begin{equation}\label{eq:var_aperture}
  \Phi_i = \sum_{j=1}^{J-1} x_{ij}\phi_j + \sum_{k=1}^{N_v} x_{ik}^{(J)}\psi_k\phi_J
\end{equation}
where the first sum represents the lamps without a variable aperture and the second sum represents the lamp with the variable aperture. The $\psi_k$, which represent the fractional opening of the variable aperture,
are constants in the interval [0,1] to be estimated. The $x_{ik}^{(J)}$ are indicator variables which take the values 0 and~1.  At most one of $x_{ik}^{(J)}$ is~1, but they can all be~0.  With $J$ lamps including a single variable aperture lamp taking one of $N_v$ levels if illuminated, there are $2^{J-1}(N_v + 1)$ unique lamp settings.

A detail of integrating spheres is left out of our model for simplicity.  An integrating sphere is a diffuse optical cavity whose throughput will change depending on how it is loaded by opening and closing shutters. Experience with the integrating sphere used in \cite{rosenberg2017preflight} indicates that this effect can be at the 0.1~\% level, and it can be accommodated by multiplying the fluxes $\phi_j$ with experimentally determined configuration factors. The experimental determinations would be made by turning on a single lamp and measuring flux changes as other shutters are opened and closed.

\subsection{Linearization Function}
\label{sec:linearization}

The model in Equation~(\ref{eq:stat_model_prob}) is a forward model in the sense that for given values of the lamp fluxes, the polynomial coefficients, and the noise, the probability distribution for the instrument signal is fixed.  However, the goal of calibration in this context is to predict the flux viewed by the detector corresponding to the instrument response.  This means that to achieve a calibration, we must invert the polynomial equation $E[n] = \alpha_0 + \sum_{m=1}^p \alpha_m P_m(s(\Phi))$, where $E[n]$ is the expected instrument reading for the flux $\Phi$. Because we are restricted to well-behaved, monotonic functions, the inverse function is also approximated by a polynomial, 
\begin{equation}
    \label{eq:inverted-poly}
    \Phi = \beta_0 + \sum_{m = 1}^p \beta_m E[n]^m .
\end{equation}
To estimate the parameters $\beta_0, \dots, \beta_p$, from $\alpha_0, \dots, \alpha_p$, a regular sequence of values is constructed over the interval $[-1, 1]$ and transformed by $s^{-1}$, the inverse of the linear transformation $s$. Denote this sequence $\widehat{\Phi}_\ell$, with $\ell = 1, \dots, L$.  The maximum likelihood estimates of $\alpha_0, \dots, \alpha_p$ are $\widehat{\alpha}_0, \dots, \widehat{\alpha}_p$, and we construct $E[\widehat{n}_\ell] = \widehat{\alpha}_0 + \sum_{m = 1}^p\widehat{\alpha}_mP_m(s(\widehat{\Phi}_\ell))$.  Finally, ordinary least squares is used to estimate $\bbeta = (\beta_0, \dots, \beta_p)$ by minimizing $\sum_{\ell=1}^{L}\left(\widehat{\Phi}_\ell - \beta_0 - \sum_{m = 1}^p \beta_m E[\widehat{n}_\ell]^m\right)^2$.  The estimate is $\widehat{\bbeta}=(\widehat{\beta}_0, \dots, \widehat{\beta}_p)$.

\subsection{Uncertainty}
\label{sec:uncertainty}

Maximizing Equation (\ref{eq:stat_model_like}) yields the MLEs for $\bphi$, $\balpha$, $\gamma$, $\sigma$, and $\bpsi$ but does not yield a quantitative uncertainty for those estimates. To get this uncertainty, we use the statistical bootstrap. This is a more efficient use of data than an alternative approach of repeating the experiment and  analysis a sufficient number of times so that uncertainties could be calculated from the distribution of results. 

Uncertainty for all parameters is quantified by a bootstrapping pairs approach \citep[see Section 7.2 and 9.5 in Ref.][for example]{efron1993introduction}. A ``pair'' is a response $n_i$ together with a configuration $\bx_i$.  The $N$ pairs of ($n_i$, $\bx_i$) are sampled at random $N$ times with no restriction on how many times a particular pair may be sampled.  Some pairs in the original dataset will appear more than once and some will not appear at all in this new bootstrap replicate. Thus, each bootstrap replicate data set will differ from the original data set, and an MLE calculated from a bootstrap replicate data set will differ from the original MLE. The distribution of replicate MLEs approximates the true sampling distribution of the MLE, and so may be used to calculate standard errors and confidence intervals. The standard error is the standard deviation of the set of bootstrap replicates, while confidence intervals are determined by computing quantiles. For example, a 95~\% confidence interval is calculated by selecting a range that includes the central 95~\% of bootstrap replicates.  

While the bootstrap algorithm accounts for sampling variability, it does not account for other sources of uncertainty like model uncertainty, i.e., the mismatch between the model and reality.  One potential mismatch is that the model in Equation~(\ref{eq:stat_model_prob}) assumes that the $\phi_j$, $j = 1,\dots, J$ are constant throughout the experiment.  If the fluxes of the lamps drift during the experiment, this is not true, but it may be accounted for in the bootstrap algorithm by also randomly perturbing $\Phi_{\rm max}$ for each bootstrap sample.  In the synthetic data in Section \ref{sec:signal_sim} two cases are considered, with lamps drifting independently of each other, and drifting together identically. These two cases represent possible extremes. Importantly, in the flux-addition method, the stimuli are applied in random order, so that systematic drift in the lamps is converted into random noise.

To account for the possibility that the lamps drift during the experiment, $\Phi_{\rm max}$ in Equation~(\ref{eq:stat_model_prob}) is perturbed by Gaussian white noise with variance $\mbox{Var}\left[\sum_j\phi_j\right]$ for each bootstrap replicate.  This can work well even if the individual lamps drift according to a distribution other than a Gaussian distribution because the assumption of Gaussian drift is placed on the sum of the lamp fluxes, which by the central limit theorem may be reasonable even if the individual lamps drift according to a different distribution. The value of $\mbox{Var}\left[\sum_j\phi_j\right]$ is an input to the analysis procedure, and it is similar to a prior distribution in Bayesian inference.  It must be based either on subject matter expertise or on data collected externally to the calibration experiment.  

\subsection{Bootstrap Sample Size Determination}

The bootstrap, as a general purpose technique for calculating standard errors and confidence intervals, is justified asymptotically.  That is, for small samples, it may not work well.  A natural question is then, how small is too small?  The authors of \cite{Stoudt2020UncertaintyEF} assert that as long as the sample size is large enough that it is likely that all bootstrap replicates will be unique, the results of the procedure should be ``reliable.''  They go on to propose sample sizes such as 20, 30, and 50, drawing from the work of \cite{chernick2011bootstrap} and \cite{hall2013bootstrap} for further support.  In the experimental scenarios considered here, very small samples sizes are not expected.  However, there are further considerations for applying a bootstrap algorithm to the cases we consider here, namely that each bootstrap replicate must sufficiently cover the whole response range $[0,\Phi_{\rm max}]$.  

Consider for pedagogy $\Phi_{\rm max} = 1$, and the following sequence of target fluxes $\Phi_i = 7.5\times10^{-5} \cdot 1.099^i, i=1,\dots,100$.  The sequence starts at about $8.2\times10^{-5}$, and ends at about 0.94.  However, only 7 values (less than 10~\%) are larger than 0.5.  The sequence was chosen for illustrative purposes, but in the Beam Conjoiner application of Sections \ref{sec:beam_conjoiner} and \ref{sec:beamcon_results} a non-uniform sequence of fluxes with more small fluxes is observed. The non-uniformity is important because in each bootstrap replicate, the probability that any one observation will be left out of a bootstrap replicate is approximately $e^{-1}$, a little more than $\frac{1}{3}$.  Thus, in the illustrative example 6 or 7 of the fluxes greater than 0.5 would be left out of about 1 out of every 100 bootstrap replicates, and 5 or more would be left out of about 1 out of every 16.  That will cause poor coverage of the flux range (0.5, 1.0), which will significantly degrade the performance of the bootstrap algorithm there.  In these cases, we recommend either changing the experiment design to avoid a non-uniform sequence of flux values or to obtain repetitions.

\subsection{Beam Conjoiner}
\label{sec:beam_conjoiner}

\begin{figure}
    \centering
    \includegraphics[width=6in]{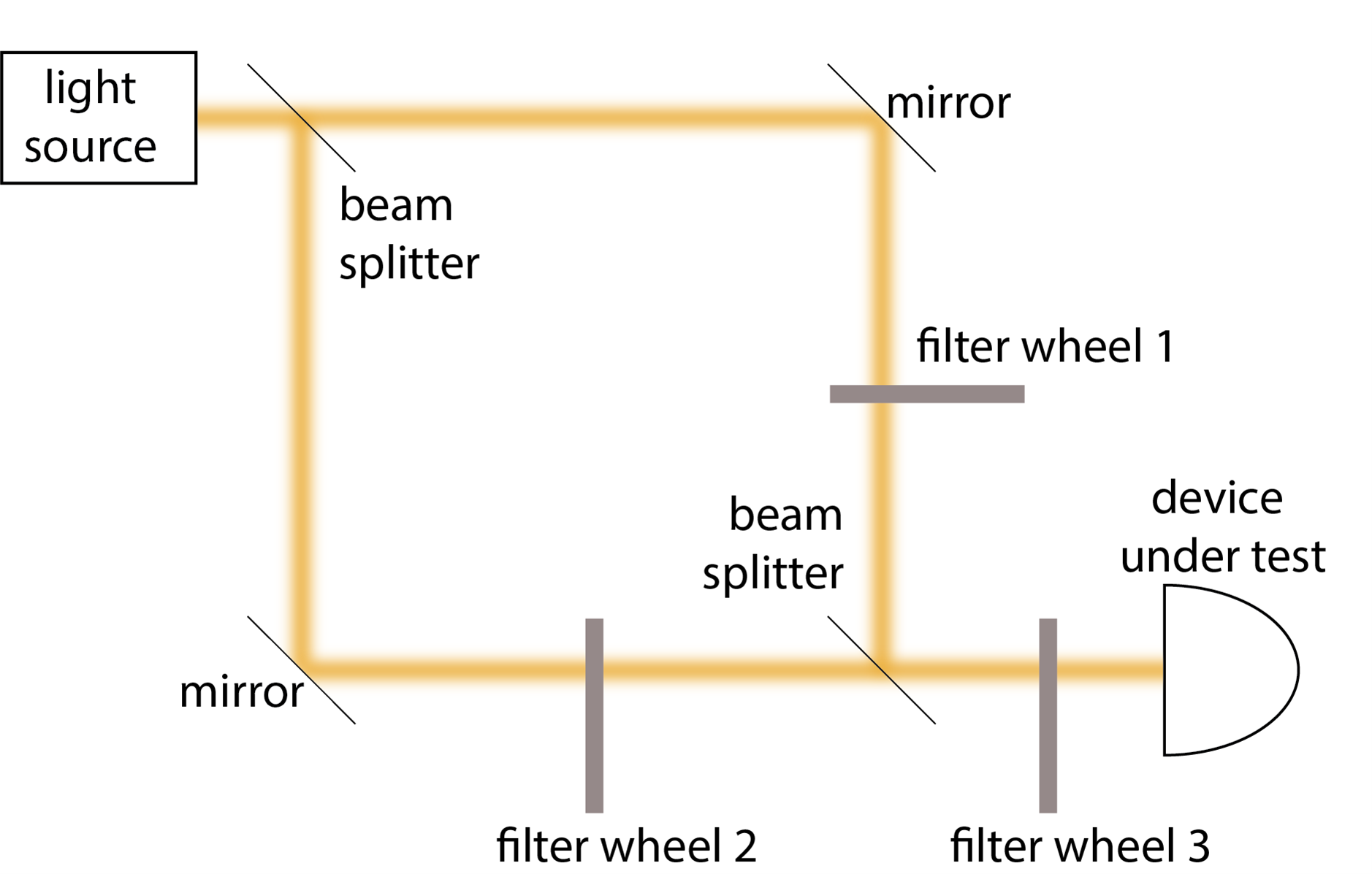}
    \caption{Schematic diagram of the Beam Conjoiner showing the primary elements.  Filter wheels are tilted so that reflections are directed into the anti-reflection coated floor of the apparatus.}
    \label{fig:beam-conjoiner-schematic}
\end{figure}

In Section \ref{sec:beamcon_results}, we describe the results of an application of the methods here to data from the NIST Beam Conjoiner, an instrument which is described in detail in \cite{thompson_beamcon_1994} and briefly here. The Beam Conjoiner comprises a light source (typically a single quartz tungsten halogen lamp) whose output is collimated and split along two different paths. Each beam passes through one neutral density filter wheel (filter 1 and filter 2) with four levels filters as well as a blocking insert. The beams are recombined and directed through a third, shared, neutral density filter wheel (filter 3) with five levels filters and a blocking insert before being viewed by the instrument. The forty filter combinations enable a dynamic range of signals of about 500 in a single measurement scan of all of the combinations.  Additional ranges are achieved using an external neutral density filter wheel which then can be used to obtain a dynamic range of 14 orders of magnitude \cite{eppeldauer1991fourteen}.  A schematic of the beam conjoiner is given in Figure \ref{fig:beam-conjoiner-schematic}.

To use the model in Equation (\ref{eq:stat_model_prob}), we make a slight adaptation for use with the beam conjoiner. There are different and equivalent ways to adapt Equation~(\ref{eq:stat_model_prob}). The simplest is to label every combination of settings for filter~1 and filter~3 as one ``lamp'' and every combination of filter~2 and filter~3 as another ``lamp''. Then constraints on the $x_{ij}$ are imposed to ensure that only physical combinations are enumerated. Thus if the $x_{ij}$ are represented as an $N\times40$ matrix, each of the first and last 20 columns would have at most a single nonzero value in each row, with each of these nonzero values representing a particular filter combination. Secondly, the sum over all lamps in the third line must be replaced with a sum over just the two brightest ``lamps.''

In our analysis of the data, we found a modification to the noise model to be necessary. We had made a simplifying assumption that instrumental noise would dominate, as represented by a single parameter $\sigma$ that is constant across a data. However, in some systems, electronic noise will dominate at low signal levels and at high signal levels the light source and photon shot noise will dominate. As a result, $\sigma$ from Equation (\ref{eq:stat_model_prob}) becomes $\sigma_i$, and for the Beam Conjoiner application, a new simplified $\sigma_i$ takes the form
\begin{equation}
    \label{eq:beam-conjoiner-noise}
    \sigma_i = \left\{
    \begin{array}{lr}
        \sigma \, \Phi_i & \mbox{if } \Phi_i > \kappa_0\Phi_{\rm max} \\
        \sigma\kappa_0\Phi_{\rm max} & \mbox{if } \Phi_i \leq \kappa_0\Phi_{\rm max}.
    \end{array}\right.
\end{equation}
In Equation~(\ref{eq:beam-conjoiner-noise}), the noise is assumed constant until the flux exceeds $\kappa_0\Phi_{\rm max}$, and after that it is assumed proportional to the flux, as would be appropriate in a case where fluctuations in the light source but not shot noise are dominant.

\subsection{Polynomial order}
\label{sec:polynomial_order}

For real-world data, it is necessary to choose the polynomial degree $p$.  The typical approach to choosing such a model parameter is to balance the trade-off between prediction variance and bias.  As $p$ increases, so too does the prediction variance, but prediction bias simultaneously decreases.

The likelihood in Equation~(\ref{eq:stat_model_like}), attempts to balance the bias-variance trade-off automatically through the parameter $\gamma$. As $\gamma$ approaches zero, $\alpha_1$ approaches $\frac{1}{2} \Phi_{\rm max}$ and the $\alpha_m$ for $m=2,\dots,p$ approach zero.  In this limit, variance is minimized (zero variance), but bias is maximized.  Conversely, as $\gamma$ approaches $\infty$, bias is minimized, but variance is maximized.  By estimating $\gamma$ along with the other parameters, the two extremes are balanced.  This implies that $p$ should be chosen to be very large to allow bias and thus mean squared error to be minimized, and that minimization is part of the estimation process.  If $p$ is too small, the bias will be artificially inflated.

Practically, however, $p$ cannot be chosen to be arbitrarily large.  A range of values of $p$ producing an optimal and stable MSE may be identified by $K$-fold cross validation (see for example Chapter 5 of \cite{james2013introduction}).  Briefly, to carry out $K$-fold cross validation, the complete data set is partitioned at random into $K$ parts. The MLEs are calculated based on $K-1$ parts, and those estimates are used to predict the instrument responses from the left out part.  An MSE is calculated using the predicted and observed instrument responses.  The process is repeated $K$ times leaving each part out of the estimation once, resulting $K$ MSEs.  

\section{Simulation Study}
\label{sec:sim_study}

The methodology described in Section \ref{sec:methodology} was tested using synthetic data under four conditions of practical interest, listed in Table \ref{tab:sim_configs}.  For each condition, 100 sets of instrument readings, the $n_i$, were generated.  For each set of $n_i$, the methodology of Section \ref{sec:methodology} was applied, and the results compared to the known truth.  The comparisons are presented in Section \ref{sec:sim_results}.  The generation of the sets of instrument readings are discussed next. 

\begin{table}
    \centering
    \begin{tabular}{r|c|c|c}
        \textbf{ID} & \textbf{identical lamps} & \textbf{linear drift} & \textbf{correlated drifts} \\
         \hline\hline 
         \textbf{1} & \cmark & 0~\% & NA \\
         \textbf{2} & \cmark &0.5~\% & \xmark \\
         \textbf{3} & \cmark &0.5~\% & \cmark \\
         \textbf{4} & 5~\% range &0.5~\% & \cmark \\

    \end{tabular}
    \caption{The four conditions considered in the simulation study. NA means ``not applicable.''}
    \label{tab:sim_configs}
\end{table}

\subsection{Signal Simulation}
\label{sec:signal_sim}

To generate one set of instrument readings $n_i$, the lamp configurations, $x_{ij}$ and $x_{ik}^{(J)}$, are first defined. The number of lamps is taken to be $J=7$.  One lamp has a variable aperture and the number of open variable aperture positions $N_v = 4$ is used. These choices are arbitrary but similar to values on existing spheres used for calibration of remote sensing instruments \cite{rosenberg_preflight_2017,Markham_OLI_2014}.  For the six lamps that do not have a variable aperture, there are $2^6 = 64$ possible off/on configurations. The variable aperture lamp can take on 5 possible states, off, on at full intensity, or on at one of the three intensities that are lower than full intensity.  This gives $64 \times 5 = 320$ possible lamp configurations.  Each simulated data set contains all 320 possible lamp configurations, plus five repetitions with all lamps off and five repetitions with all lamps on at full intensity, giving a total of $N=330$ data points in each data set.

Next, values are chosen for the $\phi_j$, $j = 1, \dots, 7$, $\psi_k$, $k = 1, \dots, 4$, and $\beta_0, \dots, \beta_p$.  In all simulations, $p=3$, $\psi_1 = 0.25$, $\psi_2 = 0.5$, $\psi_3 = 0.75$, $\beta_0 = 0.5$, $\beta_1 = 1$, $\beta_2 = 0.022$, and $\beta_3 = -0.008$.  Note that all quantities here are chosen to be dimensionless, as our procedure for recovering the inputs is used on normalized data. For simulation 1, where the lamps are identical and do not drift, $\phi_j = \frac{1}{7}$, $j=1,\dots,7$.  To represent drift in the simulations 2 through 4, each initial $\phi_j = \frac{1}{7}$ is multiplied by a random number between 0.995 and 1.005 times the number in the sequence of data points divided by $N$. In simulation 2 with ``uncorrelated drifts'', a different random number is used for each lamp in each data set. For simulations 3 and 4, which have ``correlated drifts'', the same random number is used for all lamps, with a different number used in each data set.  For simulation 4, the $\phi_j$ are randomly selected to be within approximately 2.5~\% of $\frac{1}{7}$, subject to a sum to unity constraint. This is summarized in Table \ref{tab:sim_configs}.

Given values for all of the $x_{ij}$, $x_{ik}^{(J)}$, $\phi_j$, and $\psi_k$, $\Phi_i = \sum_j^{J-1}x_{ij}\phi_j + \sum_k^{N_v}x_{ik}^{(J)}\psi_k\phi_J$ (for readability, the drifts are omitted here).  Then, $\Phi_i$ is perturbed by Gaussian white noise with standard deviation $1.1\times 10^{-4} \sqrt{\Phi_i}$ (to represent shot noise) yielding $\widetilde{\Phi}_i$. This value is for illustrative purposes. The magnitude of the shot noise selected here is in only slight conflict with the homogeneous noise model of Equation (\ref{eq:stat_model_prob}), but a real-world application may require tuning of the noise model, as is done in Section~\ref{sec:beam_conjoiner}. We found improved numerical convergence when using the homogeneous model, and it did not result in bias, as evidenced in the results section below. 

Given values for all of the $\beta_m$, $\widetilde{n}_i$ is obtained by solving the equation
\begin{equation*}
  \widetilde{\Phi}_i = \beta_0 + \sum_{m=1}^p \beta_m \widetilde{n}_i^m.
\end{equation*}
Because we restrict our analyis to well-behaved instruments with monotonic responses, there is a unique real solution.  Finally, Gaussian white noise (electrical noise) with standard deviation $10^{-3}$ is added to the $\widetilde{n}_i$ to yield $n_i$. Because of the arbitrarily chosen nonzero value for $\beta_0 = -0.5$,  $-0.5\lessapprox n_i\lessapprox 0.5$.

\subsection{Results}
\label{sec:sim_results}

To apply our method, values for $\Phi_{\rm max}$, $\tau$, and $\lambda$ must be selected.  Since for each simulation, the true values of the $\phi_j~j=1, \dots, 7$ are exactly $\frac{1}{7}$, or on average $\frac{1}{7}$ in the case of the simulations 2, 3, and 4, $\Phi_{\rm max}$ is set to unity.  The values for $\tau$ and $\lambda$ are 0.001 and 1, respectively.  For the bootstrap algorithm, it is also necessary to specify $\mbox{Var}\left[\sum_j\phi_j\right]$, as described in Section \ref{sec:uncertainty}.  For simulation 1, it is set to zero since the lamps do not not drift.  For simulations 2 and 3, it is set to 0.00055, and for simulation 4, 0.0014. The values for simulations 2, 3, and 4 are based on the known distribution of the amount of drift for the lamps.

\begin{figure}
    \centering
    \includegraphics[scale=0.65]{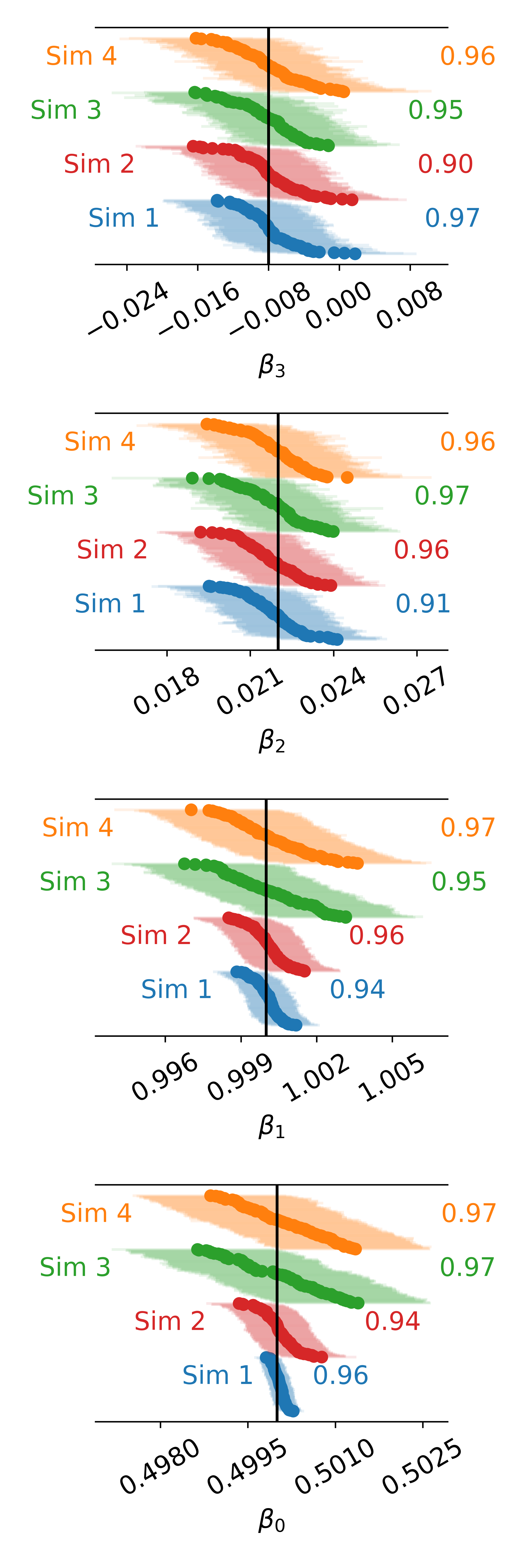}
    \caption{Results for $\beta_0$, $\beta_1$, $\beta_2$, and $\beta_3$ for 100 simulated data sets for each of the 4 simulation scenarios described in Table \protect\ref{tab:sim_configs}.  The points are the MLEs ordered from smallest to largest, and the shading represents 95~\% confidence intervals.  The vertical axis indexes the data set within a simulation, 1 to 100.  The black vertical lines are the true values, and the numbers to the right are the proportion of confidence intervals that bracket the true value.}
    \label{fig:beta_sim_results}
\end{figure}

Figure \ref{fig:beta_sim_results} shows the recovered MLEs for $\beta_0$, $\beta_1$, $\beta_2$, and $\beta_3$ for 100 simulated data sets for each of the 4 simulation scenarios described in Table \ref{tab:sim_configs}.  The points are the MLEs ordered from smallest to largest, and the shading represents 95~\% confidence intervals. The black vertical lines are the true values, and the numbers to the right are the proportion of confidence intervals that bracket the true value, the target being 0.95.  From Figure \ref{fig:beta_sim_results}, it can be seen that the true values of $\beta_0$, $\beta_1$, $\beta_2$, and $\beta_3$, lie around the center of the 100 maximum likelihood estimates for each scenario.  This implies that the estimation procedure is approximately unbiased.  To put it another way, the true values are approximately equal to the estimates, on average. For $\beta_0$ and $\beta_1$, for the 100 simulations pictured, the observed relative bias is less than 0.1~\%.  For $\beta_2$ and $\beta_3$, it is less than 1~\%.

Also from Figure \ref{fig:beta_sim_results}, it may be seen that the bootstrap procedure described in Section \ref{sec:uncertainty} achieves its nominal coverage of 95~\% for $\beta_0$, $\beta_1$, $\beta_2$, and $\beta_3$.  This claim is tenable even though none of the values on the right side of Figure \ref{fig:beta_sim_results} are exactly 0.95.  Because of the finite number (namely, 100) of data sets for each simulation, 1~of the 16 proportions in Figure~\ref{fig:beta_sim_results} is expected to lie outside of the interval $(0.91,0.99)$ even if the true coverage is exactly 95~\%, which is indeed the case.  Another important observation from Figure \ref{fig:beta_sim_results} is that for $\beta_0$ and $\beta_1$ the confidence intervals are shortest for simulation scenario 1, and longest for scenarios 3 and 4.  This occurs because of the extra uncertainty introduced into the bootstrap algorithm that is intended to account for lamp drift during the experiment.  The widths of the confidence intervals for scenario 2 are in between the two extremes of perfectly correlated drift between the lamps in scenarios 3 and 4 and no drift in scenario 1.  The widths of the confidence intervals for $\beta_2$ and $\beta_3$ do not vary systematically between the simulations, implying that uncertainty for those higher order coefficients is primarily due to sampling variability not lamp drift.

\begin{figure}
    \centering
    \includegraphics[scale=0.75]{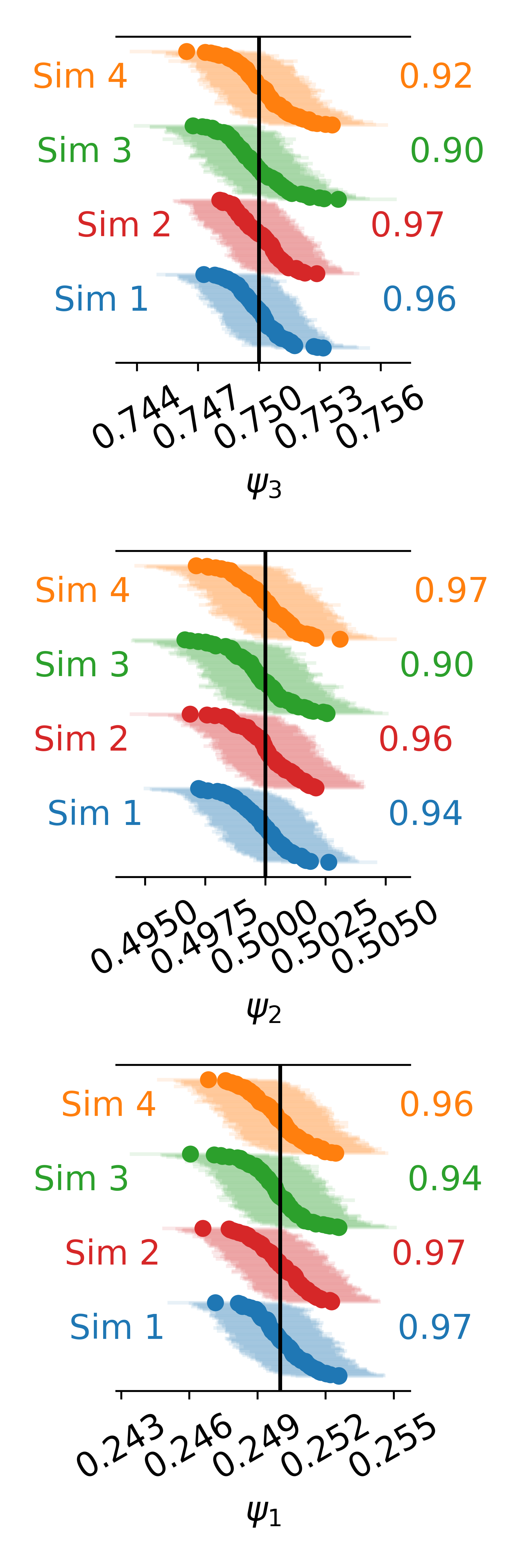}
    \caption{Results for $\psi_1$, $\psi_2$, and $\psi_3$ for 100 simulated data sets for each of the 4 simulation scenarios described in Table \protect\ref{tab:sim_configs}.  The points are the MLEs ordered from smallest to largest, and the shading represents 95~\% confidence intervals.  The vertical axis indexes the data set within a simulation, 1 to 100.  The black vertical lines are the true values, and the numbers to the right are the proportion of confidence intervals that bracket the true value, the target being 0.95.}
    \label{fig:psi_sim_results}
\end{figure}

Figure \ref{fig:psi_sim_results} shows similar results as in Figure \ref{fig:beta_sim_results}, but for $\psi_1$, $\psi_2$, and $\psi_3$.  Again, the true values are centered around the 100 maximum likelihood estimates, implying that the estimators are unbiased.  The observed relative biases are less than 0.2~\%.  Further, the coverage proportion of the confidence intervals is consistent with the nominal 95~\%.  In Figure \ref{fig:psi_sim_results}, it is also observed that the widths of the confidence intervals between simulations are similar.  As with $\beta_2$ and $\beta_3$, this implies that the dominant component of uncertainty for $\psi_1$, $\psi_2$, and $\psi_3$ is sampling variability, not lamp drift.

\begin{figure}
    \centering
    \includegraphics[scale=0.75]{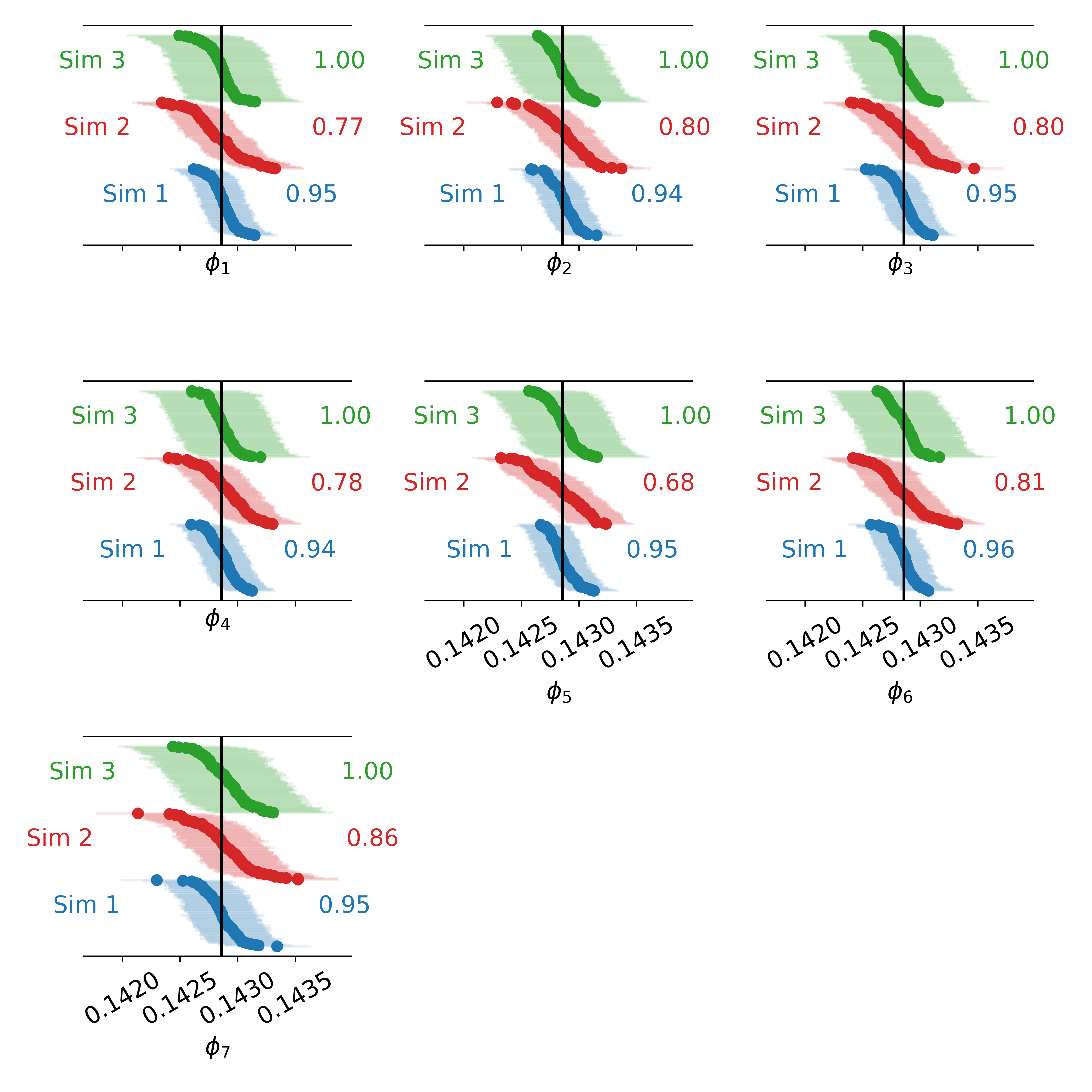}
    \caption{Results for $\phi_1,\dots,\phi_7$ for 100 simulated data sets for the first 3 of 4 simulation scenarios described in Table \protect\ref{tab:sim_configs}.  The points are the MLEs ordered from smallest to largest, and the shading represents 95~\% confidence intervals.  The vertical axis indexes the data set within a simulation, 1 to 100.  The black vertical lines are the true values, and the numbers to the right are the proportion of confidence intervals that bracket the true value, the target being 0.95.  The $4^{th}$ simulation scenario is not presented because there is not a single true value.  Tick marks refer to the same values in all graphs.}
    \label{fig:phi_sim_results}
\end{figure}

Figure \ref{fig:phi_sim_results} is similar to Figures \ref{fig:beta_sim_results} and \ref{fig:psi_sim_results}, but for $\phi_1,\dots,\phi_7$.  Simulation scenario~4 is not plotted because lamps are not assumed to be nominally identical as in scenarios 1, 2, and 3.  Again, the true average lamp fluxes are centered about the 100 maximum likelihood estimates, implying no bias. The observed relative biases are less than 0.1~\%.  On the other hand, for simulations 2 and 3, the coverage proportions of the confidence intervals are not consistent with the nominal 95~\%.  This occurs because the lamp fluxes are not exactly $\frac{1}{7}$, but instead drift around $\frac{1}{7}$ during the experiment.  The deviation of the coverage proportions from the nominal coverage for $\phi_1,\dots,\phi_7$ when the lamps drift is not concerning since the modified bootstrap algorithm accounts for the variation due to drift through $\Phi_{\rm max}$, which affects the individual $\phi_j$ only indirectly through their sum.

The simulation scenarios considered here show that the maximum likelihood estimation procedure and the non-parametric bootstrap algorithm work well for estimating nonlinear detector response as well as variable lamp intensities.  For the simulations where the lamps do not drift during the experiment, the procedures also work well for estimating the true lamp intensities.  The only cases for which the bootstrap algorithm did not convey good estimates of the uncertainty were simulation scenarios 2 and 3 (and 4, although not presented) for the true lamp intensities, $\phi_1,\dots,\phi_7$.  In those simulations, the lamps are assumed to drift during the experiment, and so in reality there is no single true value to compare the estimated lamp intensity to.  Thus, we do not find it concerning that the coverage proportions for the bootstrap confidence intervals are not consistent with the target 95~\%.  However, even in the case of drifting lamps, the maximum likelihood estimates center around $\frac{1}{7}$ because, over the 100 simulations, it is the average lamp intensity.

\section{Beam Conjoiner}
\label{sec:beamcon_results}

A dataset to evaluate the linearity of a spectroradiometer belonging to NIST was taken on the beam conjoiner. A total of four runs through each of the 150 filter wheel positions is included in our dataset. For the present work we average across a small set of pixels near the peak of the spectrum and use our methods to extract the nonlinearity. 

\begin{figure}
    \centering
    \includegraphics{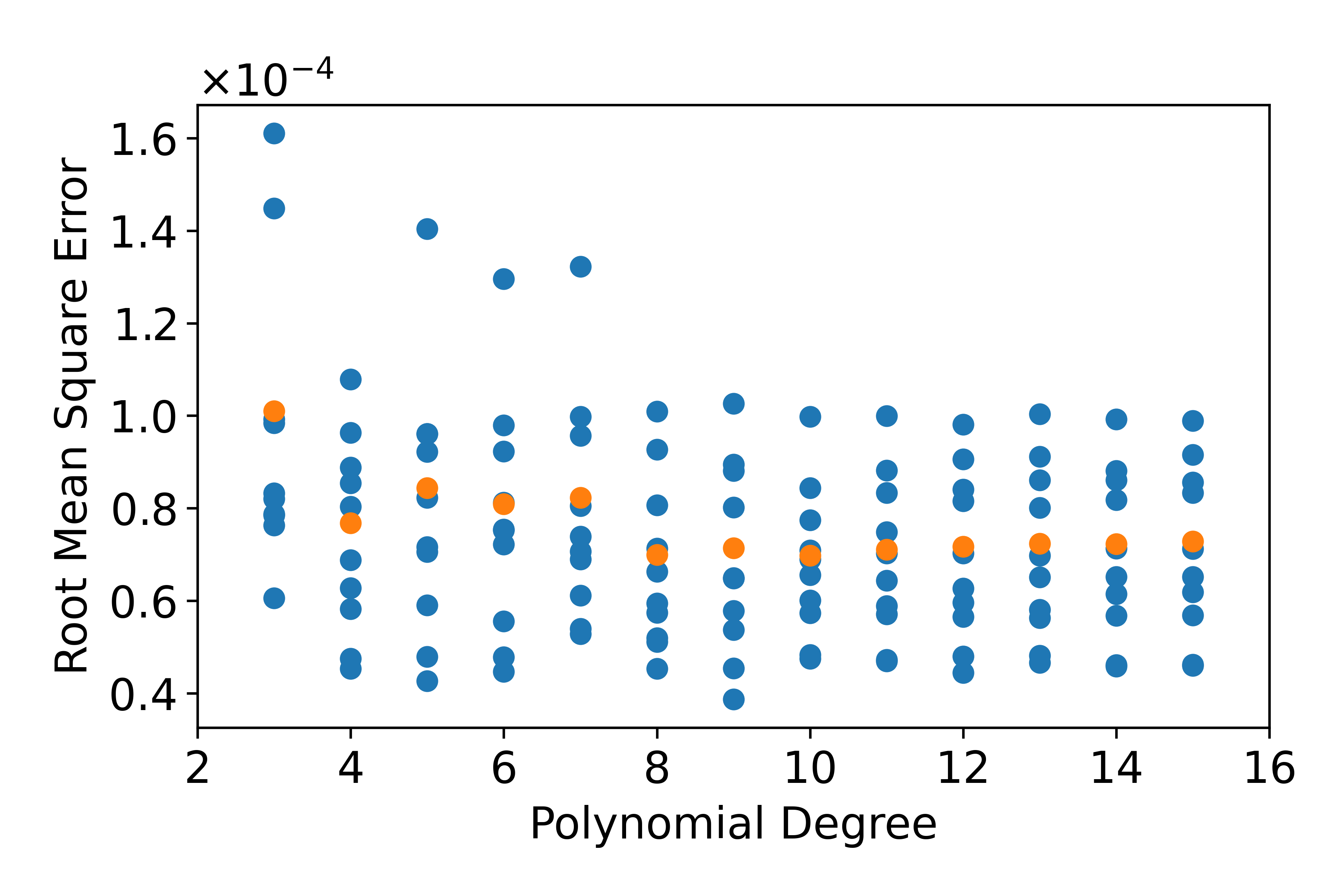}
    \caption{The 10-fold cross validation error plotted as a function of the polynomial degree $p$ for the Beam Conjoiner data discussed in Section \protect\ref{sec:beam_conjoiner}.  For each polynomial degree, the 10 estimates of root mean squared prediction error (MSE) are plotted (blue circles) as well as the square root of the average MSE (orange circles).}
    \label{fig:cv_error_beam_conjoiner}
\end{figure}

Some practical matters must be addressed before applying our methods to the Beam Conjoiner data.  First, we study the required polynomial order, per Section~\ref{sec:polynomial_order}. The results of this study are shown in in Figure \ref{fig:cv_error_beam_conjoiner}. For $8\leq p \leq 15$, the distributions of points (blue circles) and means (orange circles) are nearly identical; although, the minimum mean is achieved for $p=10$.  Based on Figure \ref{fig:cv_error_beam_conjoiner}, any $p$ in the range of $8\leq p \leq 15$ is a reasonable choice. The values of MSE in Figure \ref{fig:cv_error_beam_conjoiner} may be interpreted in the following way:  when $p$ is small, prediction bias dominates the MSE.  As $p$ increases, prediction bias decreases, and the $\gamma$ parameter in Equation~(\ref{eq:stat_model_like}) automatically balances the trade off between prediction bias and variance, which produces a range of $p$ for which the MSE is optimal and stable. 

The quantity $\kappa_0$, representing the flux level at which noise from the light source begins to dominate, from Equation (\ref{eq:beam-conjoiner-noise}) must also be chosen. In this case it was done by inspecting residuals after fitting the constant noise model, and $\kappa_0 = 0.2$ was selected. It is also necessary to select $\Phi_{\rm max}$, $\tau$, $\lambda$.  As in the simulations, $\Phi_{\rm max}$ and $\lambda$ are set to unity and $\tau=0.0001$.  Since we set $\Phi_{\rm max}=1$, the $n_i$ are also scaled to the interval $[0, 1]$ before the analysis.

\begin{figure}
    \centering
    \includegraphics{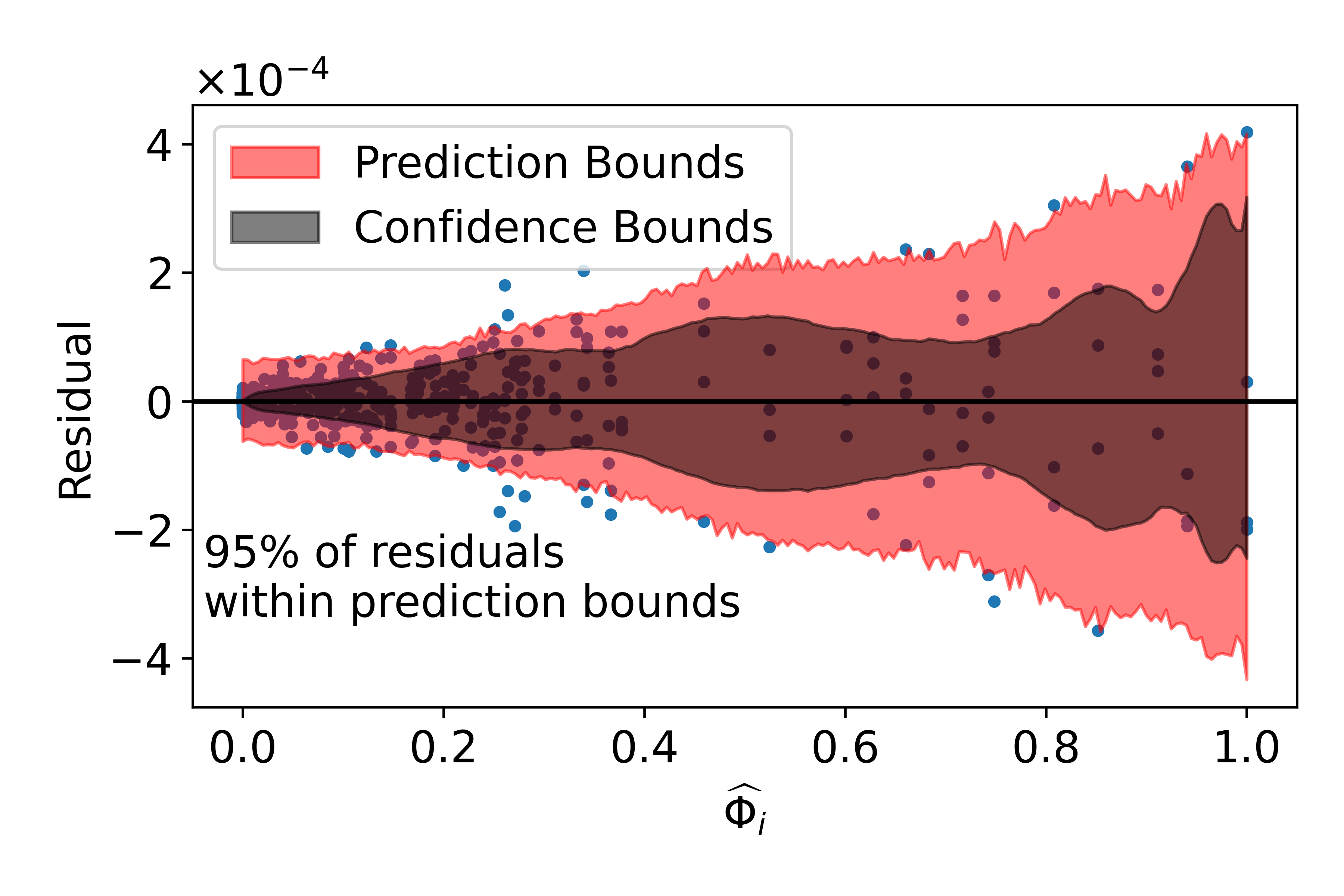}
    \caption{Residuals versus the estimated fluxes (blue points), 95~\% pointwise prediction bounds for the residuals (light red band), and 95~\% pointwise confidence bounds for the mean residual.}
    \label{fig:residual-pred-bounds}
\end{figure}

Figure \ref{fig:residual-pred-bounds} plots the residuals $n_i - \widehat{\alpha}_0 - \sum_{m=1}^p\widehat{\alpha}_mP_m(s(\widehat{\Phi}_i))$ for $p=10$ on the vertical axis versus the estimated fluxes $\widehat{\Phi}_i$ on the horizontal axis.  As noted previously, and accounted for by Equation (\ref{eq:beam-conjoiner-noise}), the increasing noise with flux is apparent in Figure \ref{fig:residual-pred-bounds}.  The residuals at each estimated flux level are symmetrically distributed around zero.  This implies that the model described in Equation~(\ref{eq:stat_model_prob}), with the modifications detailed in Section \ref{sec:beam_conjoiner}, represents the Beam Conjoiner data well.  The light red band in Figure \ref{fig:residual-pred-bounds} depicts 95~\% pointwise prediction bounds for the residuals.  The target 95~\% coverage of the prediction bounds is achieved.  This implies that the bootstrapping pairs algorithm properly accounts for uncertainty in the estimated model parameters and that the model for the noise in Equation~(\ref{eq:beam-conjoiner-noise}) is appropriate.  The light grey band in Figure \ref{fig:residual-pred-bounds} depicts 95~\% pointwise confidence bounds for the mean residual.  By construction the band contains zero, but at higher flux values and those where there are fewer observed data, the bands widen indicating more uncertainty in the average residual.

\begin{figure}
    \centering
    \includegraphics[width=6in]{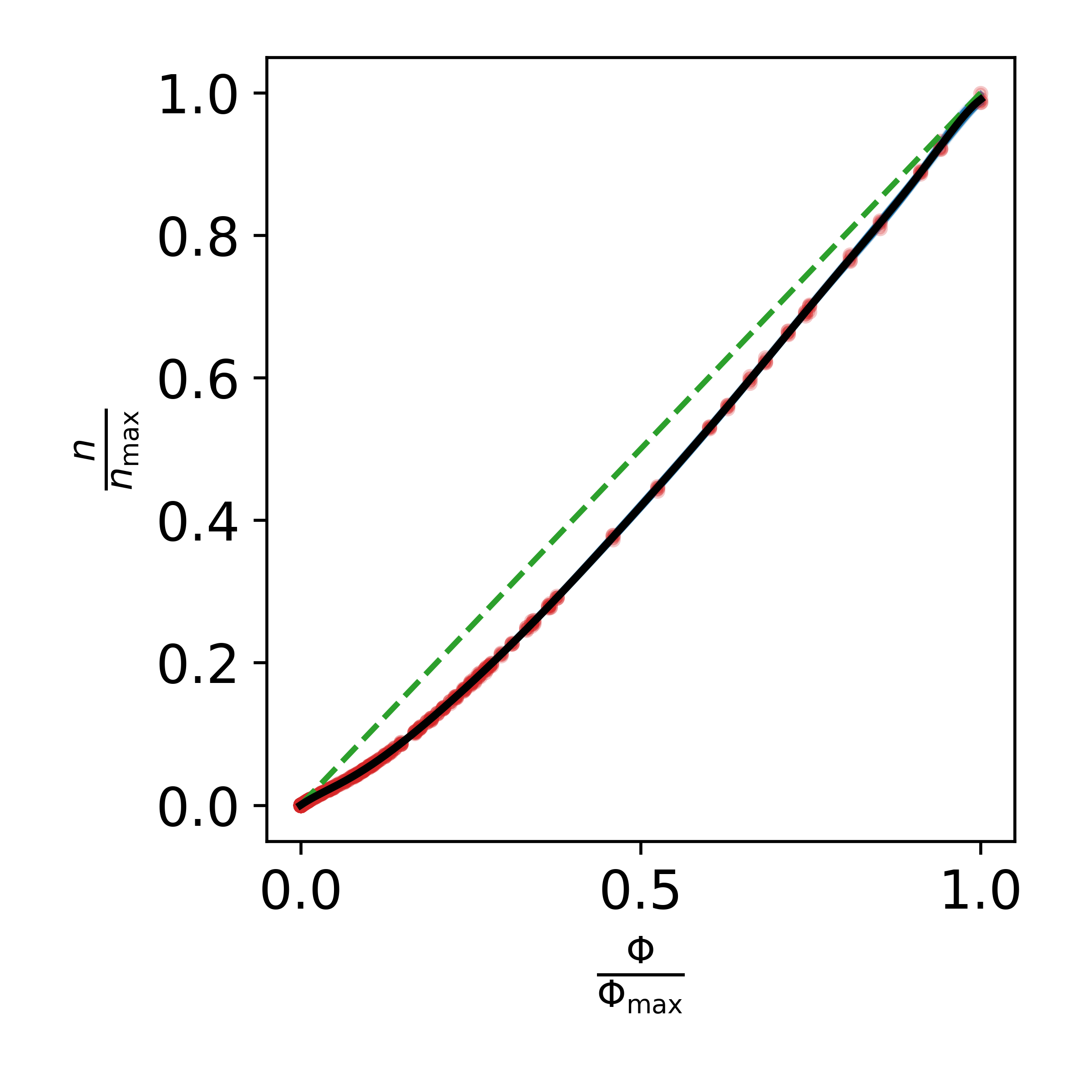}
    \caption{The horizontal axis depicts the flux, and the vertical axis depicts the instrument reading.  The black curve is the MLE of the expected instrument reading.  The red points depict the observed $n_i$.  The deviation between the expected detector response and the (green dashed) 1:1 diagonal are exaggerated by a factor of 25; otherwise, the curves would be visually indistinguishable.}
    \label{fig:beam-conjoiner-nonlin}
\end{figure}

Figure \ref{fig:beam-conjoiner-nonlin} shows the estimated nonlinear detector response where the horizontal axis represents flux and the vertical axis the detector response; the dashed green line is the diagonal, the black curve is the estimated mean detector response, and the red points are instrument readings, $n_i$.  Note that in Figure \ref{fig:beam-conjoiner-nonlin} the deviation between the diagonal and the estimated mean detector response has been exaggerated; otherwise, visually, the curves are indistinguishable.  Figure \ref{fig:beam-conjointer-nonlin-existing-vs_proposed} shows the estimated non-linear response in a different way, depicting differences between the estimated expected detector response and the diagonal (black curve), bootstrap replicates of that difference (blue curves), and differences between the the observed detector responses and the diagonal (red points); the diagonal in Figure \ref{fig:beam-conjoiner-nonlin} becomes a horizontal line at zero in Figure \ref{fig:beam-conjointer-nonlin-existing-vs_proposed}.  No exaggeration for visual effect is needed in Figure \ref{fig:beam-conjointer-nonlin-existing-vs_proposed}.  There are 982 bootstrap replicates shown because 18 of the attempted 1000 bootstrap replicates encountered failures in optimization of the likelihood function.  The Beam Conjoiner data exhibit statistically significant nonlinear behavior as seen in Figure \ref{fig:beam-conjointer-nonlin-existing-vs_proposed}, where the bootstrap replicates do not envelop zero.  The orange curve in Figure \ref{fig:beam-conjointer-nonlin-existing-vs_proposed} is discussed below in Section \ref{sec:comparison}.

\subsection{Comparison to an Existing Approach}
\label{sec:comparison}

The method for estimating $\beta_0, \dots, \beta_p$, $\phi_{11}\dots\phi_{1J_1}$, and $\phi_{21},\dots,\phi_{2J_2}$ from Ref. \cite{thompson_beamcon_1994} is based on optimizing a least-squares-like objective function $\mathcal{O}$.  
Using similar notation Equation~(\ref{eq:stat_model_like}), the objective function is
\begin{equation}
    \label{eq:ls-objective}
    \mathcal{O} =
    \left[\Phi_{\rm max} - \max_i\left\{\Phi_i\right\}\right]^2 + \sum_{i=1}^N \left[\left(\beta_0 + \sum_{m=1}^p \beta_m n_i^m\right) - \Phi_i\right]^2.
\end{equation}
The largest difference between the use of this objective function and our model is the firm statistical justification on which our model lies. While the objective function often ``works,'' it is challenging to understand where its limitations may lie. Mathematically, the objective function, unlike our model, is not derived from a generative probability model as is the likelihood function in Equation~(\ref{eq:stat_model_like}).  We consider it to be a best practice when making statistical inferences from data to begin with a generative probability model, and indeed it is a requirement in two major paradigms of statistical inference: maximum likelihood and Bayesian inference.  Further, the model presented in this paper has been adapted to deal with differing spread in the residuals at different signal levels apparent in Figure \ref{fig:residual-pred-bounds}. This adaptation has not been done with the objective function. 

\begin{figure}
    \centering
    \includegraphics{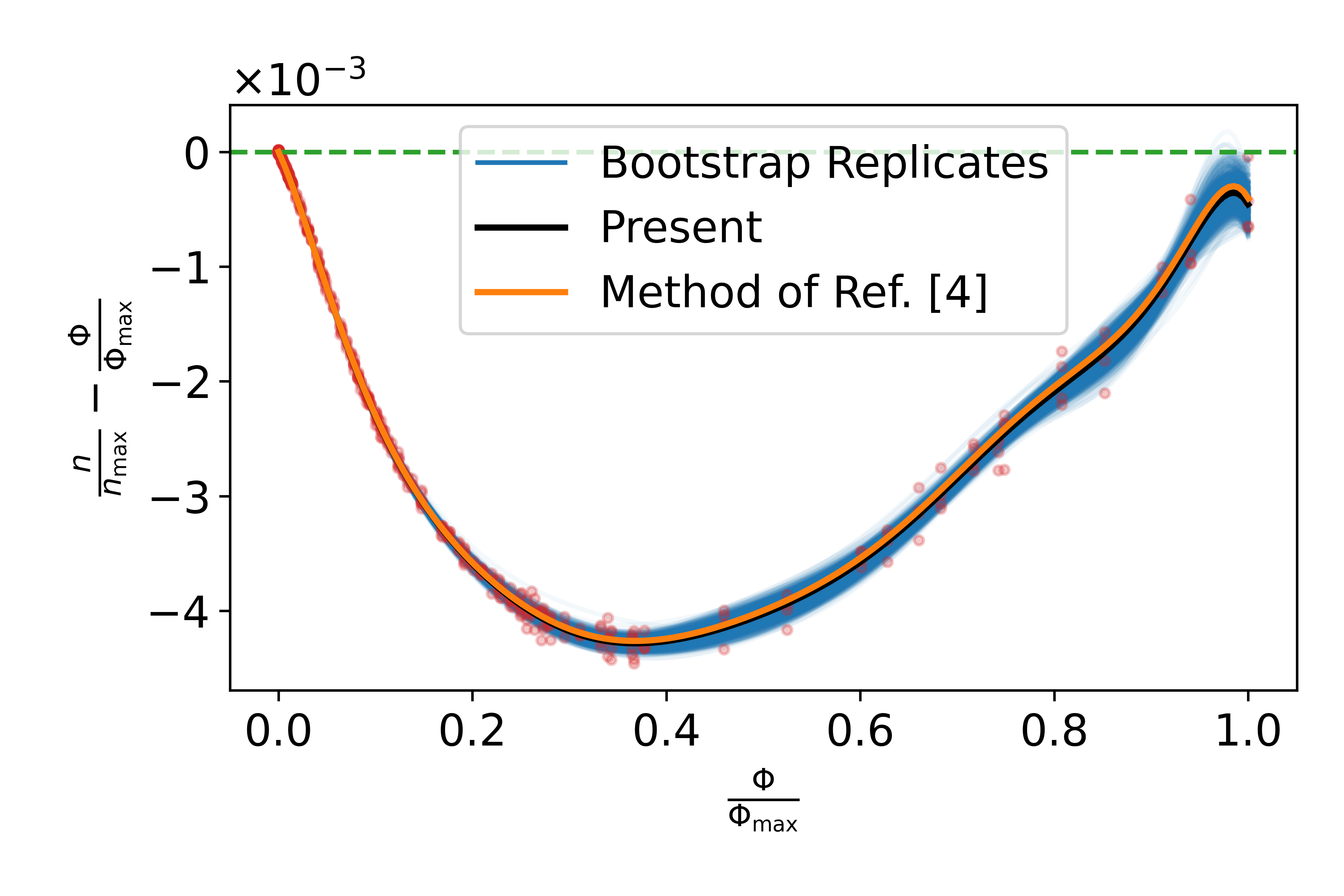}
    \caption{The horizontal axis depicts the flux, and the vertical axis depicts the instrument reading minus the flux.  The black curve is the estimated expected instrument reading minus the flux where the estimated expected instrument reading is calculated by the methods described in the present work.  The blue curves are bootstrap replicates (as described in the present work) of the estimated expected instrument reading minus the flux.  The red points depict $n_i$ minus the estimated flux for data point $i$, $\widehat{\Phi}_i$ where the $\widehat{\Phi}_i$ are calculated by the methods described in the present work.  The orange curve is an estimate of the expected instrument reading minus the flux based on Equation~(\protect\ref{eq:ls-objective}) and \protect\cite{thompson_beamcon_1994}.}
    \label{fig:beam-conjointer-nonlin-existing-vs_proposed}
\end{figure}

For comparison with our method, Figure \ref{fig:beam-conjointer-nonlin-existing-vs_proposed} includes an estimate of detector nonlinearity based on minimization of the objective function (orange curve).  The two procedures produce similar estimates.  The largest differences are found at high values of flux where there is more distinction between the approaches since the maximum likelihood approach of this work accounts for increasing variability in the detector response as flux increases.  The estimate of nonlinearity based on Equation~(\ref{eq:ls-objective}) is also  within the range of the bootstrap replicates of the estimated nonlinearity based on Equations~(\ref{eq:stat_model_prob}) and (\ref{eq:stat_model_like}) (light blue curves).  These observations provide further confidence in the proposed approach.  

To recap, both approaches produce similar estimates of the nonlinear behavior of the detector, but the approach proposed within adds flexibility and rigorous confidence and prediction intervals that may be used to express uncertainty in the results.  

\subsection{Calibration}
\label{sec:calibration}

A final step in producing a calibration using the Beam Conjoiner is to scale the results using a single, known flux level. To isolate the uncertainty in a calibration due to nonlinearity, we perform an analysis assuming we have a single known flux level $\Phi_{\rm ref}$ with zero uncertainty. Also, the average detector response, $E[n_{\rm ref}]$ is considered to have been measured a sufficient number of times to have negligible uncertainty. Extending the result to nonzero uncertainty in the $\Phi_{\rm ref}$ may be performed using Monte Carlo error propagation techniques, though we emphasize that what we show here is uncertainty due to estimation of the linearization function, which is not affected by uncertainty in the reference measurements.

\begin{figure}
    \centering
    \begin{subfigure}[b]{0.5\textwidth}
        \centering
        \includegraphics[width=\textwidth]{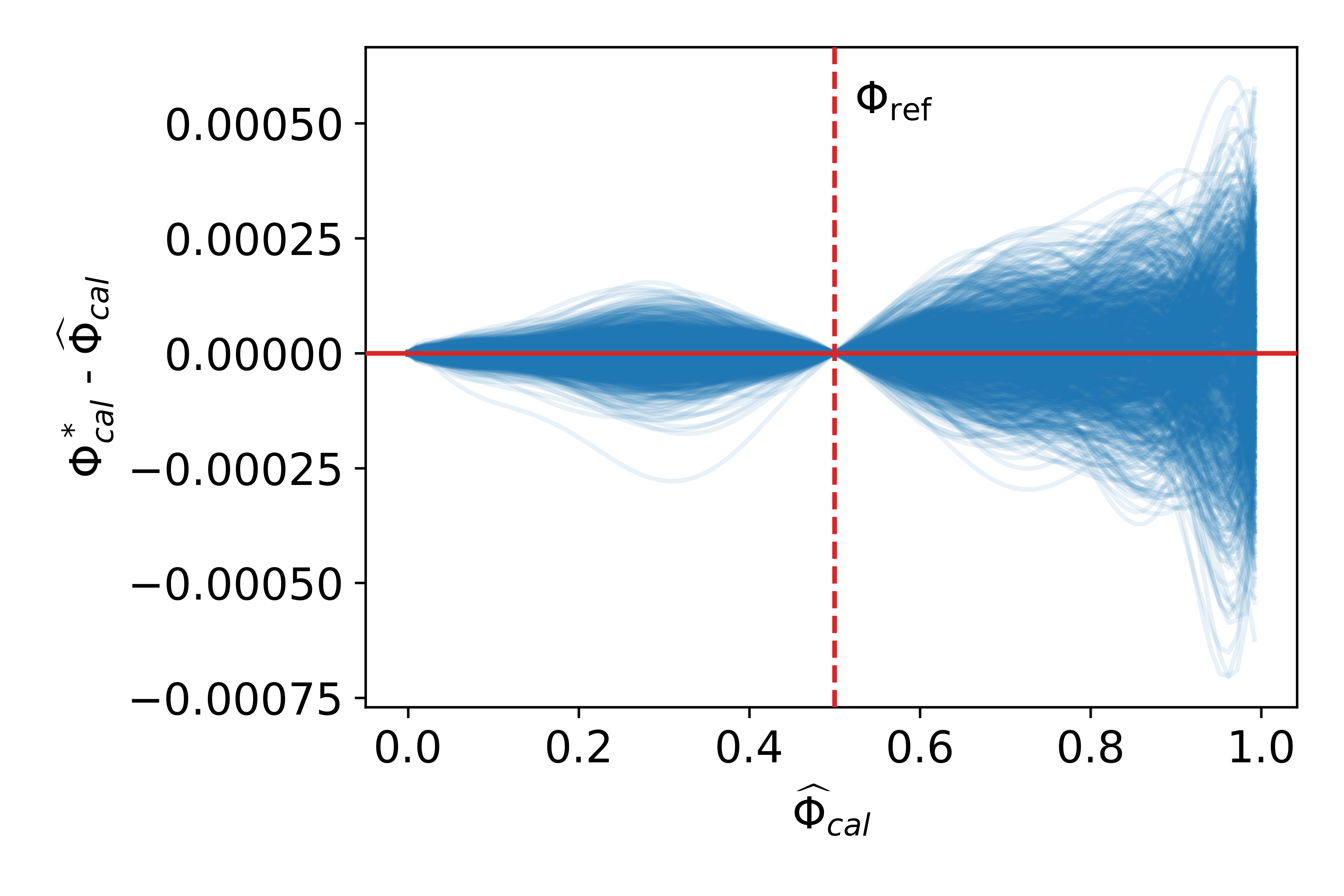}
        \caption{ }
        \label{fig:beam-conjoiner-phical-diff}
    \end{subfigure}%
    \hfill
    \begin{subfigure}[b]{0.5\textwidth}
        \centering
        \includegraphics[width=\textwidth]{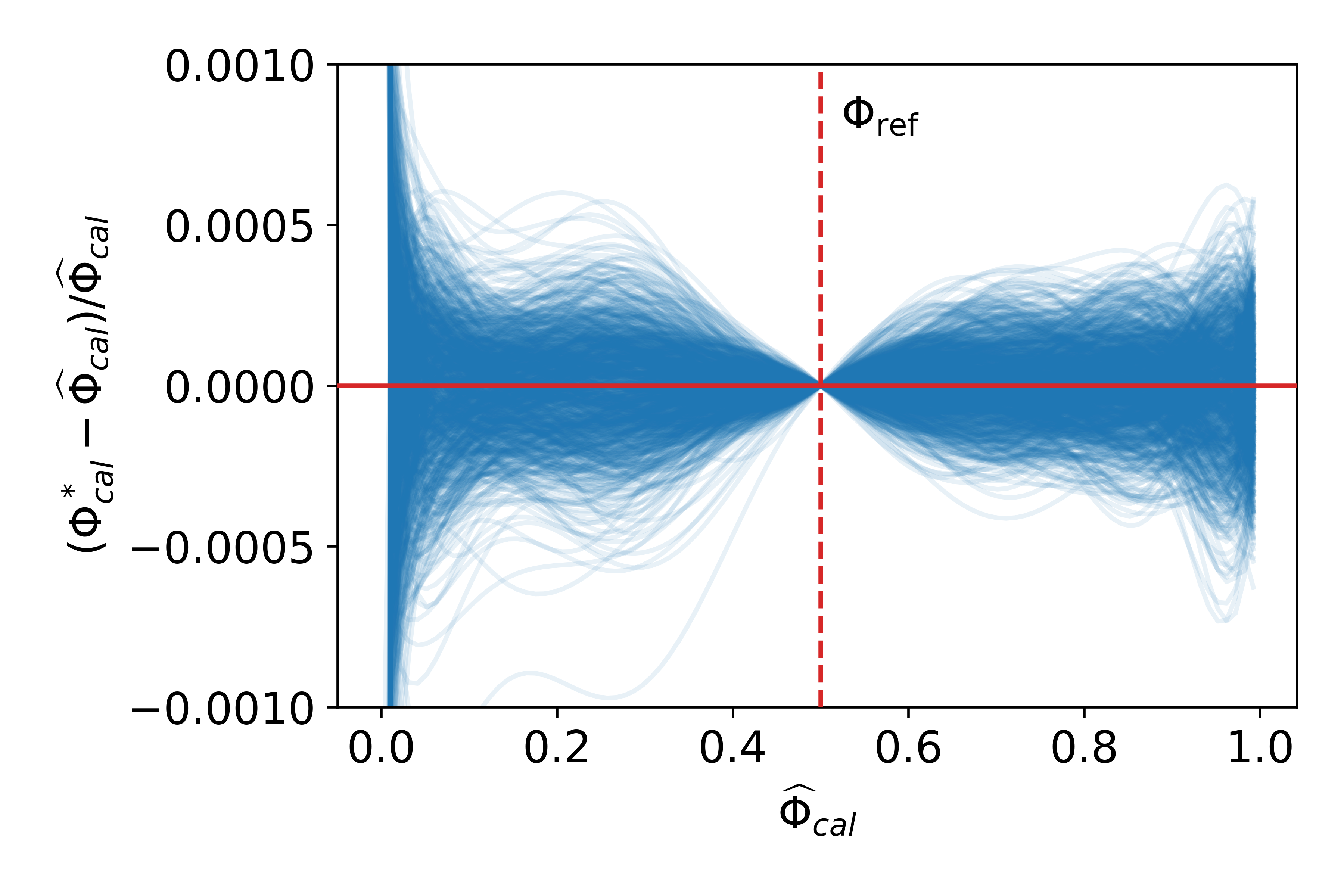}
        \caption{ }
        \label{fig:beam-conjoiner-phical-ratio}
    \end{subfigure}
    \caption{In both figures, the horizontal axis depicts the MLEs of calibrated fluxes, $\widehat{\Phi}_{\rm cal}$.  In (\protect\subref{fig:beam-conjoiner-phical-diff}) the vertical axis depicts bootstrap replicates of MLEs of calibrated fluxes minus MLEs of calibrated fluxes.  Each curve represents a single bootstrap replicate.  The vertical axis in (\protect\subref{fig:beam-conjoiner-phical-ratio}) is similar, but the relative difference is plotted.}
    \label{fig:beam-conjoiner-phical}
\end{figure}

Uncertainty estimation is done by applying the calibration to every bootstrap replicate individually, so that each predict a flux of exactly $\Phi_{\rm ref}$ when $n = E[n_{\rm ref}]$. Each bootstrap replicate represents a plausible shape for the response of the instrument, and a calibration scales and offsets these so that each give the same result at zero signal and at the single calibration level. The uncertainty attributable to empirical estimation of the linearization function is then represented by the spread in the scaled bootstrap curves as illustrated in figure \ref{fig:beam-conjoiner-phical-diff}, where we have taken $E[n_{\rm ref}] = 0.5$ and $\Phi_{\rm ref} = 0.5$, which are chosen because indicating instruments are generally designed to perform well at midrange. Figure \ref{fig:beam-conjoiner-phical-ratio} represents these data as a relative uncertainty. It may be seen directly in Figure \ref{fig:beam-conjoiner-phical-ratio}, that for most of the range of $\widehat{\Phi}_{\rm cal}$, the bootstrap replicates are within 0.025~\% of the calibrated maximum likelihood result.  

\section{Summary}
\label{sec:summary}

This article investigates the problem of quantifying uncertainty when applying the radiometry technique known as the flux-addtion method. There are four primary novel contributions:
\begin{enumerate}
    \item A generative probability model applicable to the calibration of radiometeters is introduced.  This permits the application of maximum likelihood.
    \item A non-parametric bootstrap algorithm is described for the purpose of making uncertainty statements about estimates of the parameters of the probability model.
    \item The coverage probabilities of confidence intervals derived from the bootstrap algorithm are assessed via simulation.
    \item The probability model requires the choice of a polynomial degree.  A cross-validation-based strategy is proposed to help make that choice.
\end{enumerate}

The probabilistic model and bootstrap algorithm were tested in a simulation study.  In the simulation study, the MLEs for the polynomial coefficients of the linearizing function were found to be approximately unbiased, meaning when estimates are averaged across many data sets, the average value is approximately the true value.  For the polynomial coefficients defining the linearization function, the relative bias observed in the simulation study was less than 1~\%. The simulation study also shows that bootstrap percentile confidence intervals for the same polynomial coefficients are consistent with their nominal 95~\% coverage.  In 100 simulations, the observed coverage proportions ranged from 90~\% to 97~\%.  Similar results were obtained for the other model parameters, e.g., fractional flux factors for variable aperture lamps.  The coverage results held for some model parameters, most notably $\bbeta$, even when the model assumption of constant lamp fluxes, over the duration of the experiment, was relaxed. We note that to use the methodology on a real satellite instrument would require careful planning to ensure that the measurements and the model are well-matched. 

The probabilistic model and bootstrap algorithm were also tested on an experimental data set from the NIST Beam Conjoiner. The model was deemed to perform well for these data because pointwise 95~\% prediction intervals for the residuals enveloped exactly 95~\% of them.  The probabilistic model was also compared to a conventional technique.  The two approaches matched well for estimating the linearization function. 

After a further calibration step, for the Beam Conjoiner data set, the component of uncertainty in a calibrated value, attributable to estimation of the linearizing function, was found to be less than 0.025~\% for most of the range of the instrument.  The method provides unbiased estimates for nonlinear corrections required in high-precision calibrations along with accurate uncertainty statements.

\section*{Acknowledgements}
The authors
thank Robert Rosenberg of NASA Jet Propulsion Laboratory for helpful conversations and feedback.

\section*{Disclaimers}
Certain commercial products or company names are identified here to describe our study adequately. Such identification is not intended to imply recommendation or endorsement by the National Institute of Standards and Technology, nor is it intended to imply that the products or names identified are necessarily the best available for the purpose.
The manuscript is a product of the United States Government and not subject to copyright in the United States.

\goodbreak

\appendix

\section{Table of Symbols}
\label{sec:symbols}

{\centering
\begin{longtable}{|lp{\linewidth}|}
    \caption{Table of symbols.  If a symbol appears in the \textit{Integrating Sphere} section, it is not repeated in the following sections, e.g., $\Phi_i$. With the exception of symbols appearing in the calibration section, accented symbols, such as $\widehat{\alpha}_0$, are not included because the accent does not change the underlying meaning of the symbol.  For $\widehat{\alpha}_0$, the accent designates that it is the MLE of the unaccented symbol.  The meaning of all accents are discussed at first use.  In the calibration section, some symbols are originally defined with accents.}
    \label{tab:symbol_defs} \\
    \hline
    \multicolumn{2}{|c|}{Integrating Sphere} \\
    \hline \hline
     $n_i$ & Instrument reading for data point $i=1,\dots,N$. \\
     $n$ & General symbol denoting an instrument reading. \\
     $N$ & Number of instrument readings. \\
     $\alpha_m$ & Legendre polynomial coefficients; $m=1,\dots,p$. \\
     $p$ & Maximum polynomial degree. \\
     $P_m(\cdot)$ & Legendre polynomial of order $m$. \\
     $s(\cdot)$ & Function that shifts and scales inputs to the interval $[-1, 1]$. \\
     $\Phi_i$ & Stimulus or flux for data point $i$. \\
     $\Phi$ & General symbol denoting a stimulus or flux. \\
     $\sigma$ & Standard deviation of noise. \\
     $\phi_j$ & Flux for lamp $j=1,\dots,J$. \\
     $x_{ij}$ & Indication of lamp $j$ being 0 (off) or 1 (on) for data point $i$. \\
     $\Phi_{\rm max}$ & Estimate of maximum flux. \\
     $\tau$ & Uncertainty in estimated maximum flux expressed as a standard deviation. \\
     $\gamma$ & Parameter used to force the Legendre polynomial coefficients. \\
     $\lambda$ & Prior expected value of the parameter used to force the polynomial coefficients. \\
     $\psi_k$ & Scale in $[0,1]$ for variable aperture setting $k=1,\dots,N_v$. \\
     $x_{ik}^{(J)}$ & Indication of variable aperture state $k$ on lamp $J$ for data point $i$. \\
     $N_v$ & Number of variable aperture settings. \\
     $\beta_m$ & Polynomial coefficients defining the linearization function $h^{-1}$.\\
     \hline
     \hline
     \multicolumn{2}{|c|}{Beam Conjoiner} \\
     \hline \hline
     $\sigma_i$ & Standard deviation of  noise for data point $i$. \\
     $\kappa_0$ & The cutoff below which noise is assumed constant is $\kappa_0\Phi_{\rm max}$. \\
     $\sigma$ & The standard deviation of the noise is assumed to be proportional to the flux with proportionality constant $\sigma$. \\
     \hline
     \hline
     \multicolumn{2}{|c|}{Calibration} \\
     \hline \hline
     $\Phi_{\rm ref}$ & Reference flux. \\
     $E[n_{\rm ref}]$ & Expected instrument response at $\Phi_{\rm ref}$. \\
     $\widehat{\Phi}_{\rm cal}$, $\Phi_{\rm cal}^*$ & Calibrated flux. \\
     \hline
\end{longtable}
}

\bibliographystyle{iopart-num}
\bibliography{main}

\providecommand{\newblock}{}
\begin{thebibliography}{10}
\expandafter\ifx\csname url\endcsname\relax
  \def\url#1{{\tt #1}}\fi
\expandafter\ifx\csname urlprefix\endcsname\relax\def\urlprefix{URL }\fi
\providecommand{\eprint}[2][]{\url{#2}}

\bibitem{vim2012}
\mbox{Joint Committee for Guides in Metrology (JCGM)} 2012 {\em International
  vocabulary of metrology – Basic and general concepts and associated terms
  (VIM)\/} (International Bureau of Weights and Measures (BIPM))
  \urlprefix\url{https://www.bipm.org/documents/20126/2071204/JCGM\_200\_2012.pdf}

\bibitem{white_general_2008}
White D~R, Clarkson M~T, Saunders P and Yoon H~W 2008 {\em Metrologia\/} {\bf
  45} 199--210 ISSN 0026-1394
  \urlprefix\url{https://doi.org/10.1088\%2F0026-1394\%2F45\%2F2\%2F009}

\bibitem{efron1993introduction}
Efron B and Tibshirani R~J 1993 {\em An Introduction to the Bootstrap\/} (CRC
  press)

\bibitem{thompson_beamcon_1994}
Thompson A~K and Chen H~M 1994 {\em J. Res. Natl. Inst. Stand. and Technol.\/}
  {\bf 99} 751--755

\bibitem{shin2013high}
Shin D~J, Park S, Jeong K~L, Park S~N and Lee D~H 2013 {\em Metrologia\/} {\bf
  51} 25--32

\bibitem{LegendreFunction}
Legendre and related functions, {Chap}.~14 in the \textit{NIST Digital Library
  of Mathematical Functions} http://dlmf.nist.gov/, Release 1.1.3 of
  2021-09-15, {F}.~W.~J. Olver, A.~B. {Olde Daalhuis}, D.~W. Lozier, B.~I.
  Schneider, R.~F. Boisvert, C.~W. Clark, B.~R. Miller, B.~V. Saunders, H.~S.
  Cohl, and M.~A. McClain, eds. \urlprefix\url{http://dlmf.nist.gov/14}

\bibitem{wood2015core}
Wood S 2015 {\em Core Statistics\/} (Cambridge University Press) ISBN
  9781107071056

\bibitem{hoerl1970ridge}
Hoerl A~E and Kennard R~W 1970 {\em Technometrics\/} {\bf 12} 55--67

\bibitem{tibshirani1996regression}
Tibshirani R 1996 {\em Journal of the Royal Statistical Society: Series B
  (Methodological)\/} {\bf 58} 267--288

\bibitem{gelman2013bayesian}
Gelman A, Carlin J, Stern H, Dunson D, Vehtari A and Rubin D 2013 {\em Bayesian
  Data Analysis, Third Edition\/} Chapman \& Hall/CRC Texts in Statistical
  Science (Taylor \& Francis) ISBN 9781439840955

\bibitem{park2008bayesian}
Park T and Casella G 2008 {\em Journal of the American Statistical
  Association\/} {\bf 103} 681--686

\bibitem{rosenberg_preflight_2017}
Rosenberg R, Maxwell S, Johnson B~C, Chapsky L, Lee R~A~M and Pollock R 2017
  {\em IEEE Transactions on Geoscience and Remote Sensing\/} {\bf 55}
  1994--2006 ISSN 0196-2892

\bibitem{rosenberg2017preflight}
Rosenberg R, Maxwell S, Johnson B~C, Chapsky L, Lee R~A and Pollock R 2017 {\em
  IEEE Transactions on Geoscience and Remote Sensing\/} {\bf 55} 1994--2006

\bibitem{Stoudt2020UncertaintyEF}
Stoudt S, Pintar A~L and Possolo A 2020 {\em Metrologia\/} {\bf 58}

\bibitem{chernick2011bootstrap}
Chernick M~R 2011 {\em Bootstrap methods: A guide for practitioners and
  researchers\/} (John Wiley \& Sons)

\bibitem{hall2013bootstrap}
Hall P 2013 {\em The bootstrap and Edgeworth expansion\/} (Springer Science \&
  Business Media)

\bibitem{eppeldauer1991fourteen}
Eppeldauer G and Hardis J~E 1991 {\em Applied Optics\/} {\bf 30} 3091--3099

\bibitem{james2013introduction}
James G, Witten D, Hastie T and Tibshirani R 2013 {\em An Introduction to
  Statistical Learning: with Applications in R\/} Springer Texts in Statistics
  (Springer New York) ISBN 9781461471387

\bibitem{Markham_OLI_2014}
Markham B, Barsi J, Kvaran G, Ong L, Kaita E, Biggar S, Czapla-Myers J, Mishra
  N and Helder D 2014 {\em Remote Sensing\/} {\bf 6} 12275--12308 ISSN
  2072-4292 \urlprefix\url{https://www.mdpi.com/2072-4292/6/12/12275}

\end{thebibliography}

\end{document}